# Observation of Large and All-Season Ozone Losses over the Tropics

Qing-Bin Lu*

Department of Physics and Astronomy and Departments of Biology and Chemistry, University of Waterloo, 200 University Avenue West, Waterloo, Ontario, Canada

*Corresponding author (Email: qblu@uwaterloo.ca)

**Abstract**: This paper reveals a large and all-season ozone hole in the lower stratosphere over the tropics (30°N-30°S) since the 1980s, where an $O_3$ hole is defined as an area of $O_3$ loss larger than 25% compared with the undisturbed atmosphere. The depth of this tropical $O_3$ hole is comparable to that of the well-known springtime Antarctic $O_3$ hole, whereas its area is about seven times that of the latter. Similar to the Antarctic $O_3$ hole, approximately 80% of the normal $O_3$ value is depleted at the center of the tropical $O_3$ hole. The results strongly indicate that both Antarctic and tropical $O_3$ holes must arise from an identical physical mechanism, for which the cosmic-ray-driven electron reaction (CRE) model shows good agreements with observations. The whole-year large tropical $O_3$ hole could cause a serious global concern as it can lead to increases in ground-level ultraviolet radiation and affect 50% of Earth's surface area, home to approximately 50% of the world's population. Moreover, the presence of the tropical and polar $O_3$ holes is equivalent to the formation of three 'temperature holes' observed in the stratosphere. These findings will have significances in understanding planetary physics, ozone depletion, climate change, and human health.

## 1. INTRODUCTION

The successful execution of the Montreal Protocol and its revisions has led to the declining total level of tropospheric ozone-depleting substances (ODSs) (mainly chlorofluorocarbons-- CFCs) since around 1994. Yet, the $O_3$ hole over the Arctic in 2020 set a biggest record,[1] while the $O_3$ holes over the Antarctic in 2020 and 2021 were among the largest, deepest and most persistent ones. As a matter of fact, these observations were not expected from photochemical models[2] but precisely predicted by the cosmic-ray-driven electron-induced-reaction (CRE) model.[3-9] There are also significant research interests in electron-induced reactions of halogen-containing molecules including CFCs in the gas phase and clusters and on ice surfaces[10-15] and in the effects of cosmic rays (CRs) on Earth's clouds, climate and space weather[16-20].

This study was motivated by the following observations. First, the author[9] has recently shown that the CRE-initiated $O_3$-depleting reaction completely destroys or even overkills the $O_3$ layer in the lower Antarctic stratospheric region at altitudes of 13.5-17.5 km corresponding to the CR ionization maximum. Second, CRs have a latitude effect due to the geomagnetic field, which leads to a maximum intensity in the polar stratosphere and a lower intensity by about 50% over the tropics, whereas CFCs have a spatial distribution anti-correlated with the intensity of CRs.[4-8] Third, critically there exist all-season low-temperature cyclones centered at the altitude of around 15 km over the tropics (30°N-30°S), in which the CR ionization intensity peaks and the temperatures are low at 190-200 K, as will be shown by observed data in this study. The latter is comparable to that in the polar stratospheric vortex over Antarctica in winter and cold enough for forming tropical stratospheric clouds (TSCs), similar to polar stratospheric clouds (PSCs) that are required for surface reactions leading to an $O_3$ hole.[21] It is important to note that the composition and climate of the lower stratosphere over the tropics is very different from those over the polar regions. This should give rise to a unique active halogen evolution for the tropical lower stratosphere: with the constant presence of TSCs, there are no stable chlorine/bromine reservoirs (HCl and $ClONO_2$) as they are rapidly destroyed on the surfaces of TSCs, while with the constant presence of intense sunlight, no ClO dimer ($Cl_2O_2$) can exist and $O_3$-depleting reaction cycles are much more effective. Therefore, halogen-catalyzed reactions are much more efficient for tropical $O_3$ destruction, and even a low level of active halogen can cause significant ozone depletion over the tropics in all seasons. The $O_3$ holes over Antarctica[22] and the Arctic[23] have been well observed, whereas no $O_3$ hole over the tropics has been reported.[2] Nevertheless, there have been reports of significant ozone losses over the tropics, implying the possibility of a tropical ozone hole. For example, Bian et al.[24] and Chen et al.[25] reported observations of an "ozone mini hole" over the Tibetan Plateau (29.6° N, 91.1° E) at times with colder winters in the 1990s-2000s, which is located at the edge of the tropics. Polvani et al.[26] observed significant $O_3$ losses at the altitude of 18.5 km or 67/68 hPa over the tropics (30°S-30°N) from 1979 to 1997 from three independent datasets. Newton et al.[27] observed ozone-poor air (with very low $O_3$ concentrations) in the tropical tropopause layer over the west Pacific by aircraft measurements from the three experiment campaigns based in Guam in January–March 2014. Basing on the above reasoning, the author hypothesized that there likely exists a large $O_3$ hole over the tropics, which should be comparable to the well-known Antarctic $O_3$ hole in depth and much larger than the latter in area. Therefore, this article is devoted to a search for the tropical $O_3$ hole.

Despite the rationales given above, the search for the tropical $O_3$ hole is challenging. This is due to some intrinsic challenges. First, unlike polar $O_3$ holes that are seasonal and mainly appear in spring, a tropical $O_3$ hole if exists is essentially unchanged across the seasons and is therefore invisible in original observed data. Second, unlike in polar regions, the tropical $O_3$ layer mainly lies in the middle stratosphere (≥25 km) and only a relatively small percentage (25-30%) of total $O_3$ is distributed in the lower tropical stratosphere (at altitudes of 10-25 km). Thus, the amplitude of $O_3$ depletion in the lower tropical stratosphere is not explicitly reflected in measured total $O_3$ data. An associated fact that led to no previous observations of any $O_3$ hole over the tropics is the conventional definition of an $O_3$ hole, which is defined as an area in which total $O_3$ values drop below the historical threshold of 220 Dobson Units (DU) (the total $O_3$ value measured by the NASA TOMS satellite in 1979). With this definition, no tropical $O_3$ holes could be observed in satellite images of total $O_3$ even if a large percentage up to ~65% of the $O_3$ amount in the lower tropical stratosphere is depleted. Moreover, much attention was drawn to the studies of $O_3$ depletion over the polar regions, as no $O_3$ hole was expected to appear over the tropics from current photochemical models. Fortunately, ground-based measurements since the 1960s combined with satellite measurements since 1979 of $O_3$, CFCs and temperature in the stratosphere still provide sufficient data to examine the above hypothesis. Here an $O_3$ hole is defined as an area with $O_3$ loss in percent larger than 25%, with respect to the undisturbed $O_3$ value when there were no significant CFCs in the stratosphere (~ in the 1960s). Despite its difference from the conventional definition of an $O_3$ hole, this new definition is supported by the observed $O_3$ loss (~25% in spring) over the Arctic (shown later) where and when an $O_3$ hole was reported. In other words, an area with $O_3$ loss by 25% is approximately the threshold of observing an Arctic $O_3$ hole.

## 2. DATA AND METHODS

A standard troposphere-stratosphere $O_3$ climatology in altitude relative to the ground level in the Trajectory-mapped Ozonesonde dataset for the Stratosphere and Troposphere (TOST) obtained from the World Ozone and Ultraviolet Radiation Data Centre (WOUDC), latitude-altitude CFC distribution, time-series total ozone, Ozonesonde and Umkehr datasets, and lower stratospheric temperature climatology from multiple ground-based or satellite measurements are used for this study. TOST is a global 3-D (latitude, longitude, altitude) climatology of tropospheric and stratospheric $O_3$, derived from the WOUDC global $O_3$ sounding record by trajectory mapping and uses approximately 77,000 ozonesonde profiles from some 116 stations worldwide since 1965, with details given by Liu et al.[28] The long-term zonal mean latitude-altitude distribution of the $O_3$ climatology measured over the past five decades (1960s-2010s) is shown in Fig. S1 in Supplementary Material (SM).

To reveal the tropical $O_3$ hole, we show the differences of the $O_3$ climatology in 1970s, 1980s, 1990s, 2000s and 2010s with respect to that in 1960s. It is critical to note that the distribution of $O_3$ over the tropics is highly ununiform in altitude. Thus, the differences in zonal mean latitude-altitude distribution of the $O_3$ climatology do not simply reflect the value of $O_3$ depletion by a physical mechanism, and there are larger $O_3$ losses in absolute values at higher altitudes. Thus, we plot the relative $O_3$ changes in percent (%) in 1970s-2010s with respect to that in the 1960s to reveal the tropical ozone hole. Furthermore, to confirm the robustness of the observation, we also show the seasonal $O_3$ changes in percent in 2000s *with respect to those in the 1980s* since the latter ozone data should be more reliably measured and more abundant than those in the 1960s. Moreover, the observed results of the present study are also compared with those of Polvani et al.[26] using TOST[28] and other datasets from combined ground-based and satellite measurements[29, 30].

In addition, the relative $O_3$ change in percent in an $O_3$ hole is given by the equation derived from the CRE mechanism:[6-8]

$$\Delta[O_3]/[O_3]_0 = k[C]I^2, \quad (1)$$

where $[O_3]_0$ is the normal $O_3$ amount when there were no significant stratospheric effects of CFCs (i.e., $[O_3]_0 \approx [O_3]$ in the 1960s), $[C]$ is the equivalent effective chlorine (EECl) in the stratosphere, $I$ is the CR intensity, and $k$ is a fitting constant determined by the best fit to past measured $O_3$ data. Thus, we also use Eq. (1) to analyze the observed ozone and temperature data in the stratosphere.

## 3. RESULTS

### 3.1 Ozone climatology

The large Antarctic $O_3$ hole is well known to appear in the spring, while the hypothesized tropical $O_3$ hole is expected to appear in all four seasons. Thus, we first show the observed results for the Antarctic spring season (September, October and November—SON) in order to compare the tropic $O_3$ hole if exists with the well-known Antarctic $O_3$ hole. As shown in Fig. S2, the thus plotted result shows that from the 1960s to the 1970s, there were almost no ozone holes but small ozone decreases at 10-20 km over the Antarctica and at 16-25 km over the tropics and there was a tropospheric-stratospheric region of significant ozone *increases* at altitudes below 16 km and latitudes of 10°N-40°S, which was likely caused by volcanic effects. These small $O_3$ decreases in the Antarctic and tropical lower stratospheres are expected, as it is known that the reaction of stratospheric CFCs remained insignificant up to the 1970s. Despite that there were arguably more sparse measured data available for constructing the TOST data for the 1960s, the result in Fig. S2 shows the validity of the 1960s' TOST data.

As plotted in Fig. 1, in contrast to Fig. S2, the observed results for the 1980s-2010s with respect to that of the 1960s show clearly a deep and large tropical $O_3$ hole with significant $O_3$ loss since the 1980s, along with the well-known Antarctic $O_3$ hole. The Antarctic $O_3$ hole reached the largest and deepest in the 1990s, while the tropic hole continued to grow and deepen in the 1990s and reached its maximum in the 2000s. For both largest and deepest Antarctic (1990s) and tropical (2000s) $O_3$ holes, the same maximum (decadal averaged) $O_3$ loss of ~80% at the hole centers was observed. However, it is worthwhile noting that the geometric area of the tropical $O_3$ hole at latitudes of 30°S-30°N is approximately 7.5 times that of the Antarctic hole at 60°-90°S. Both holes in the 2010s were slightly smaller than those in the 2000s.

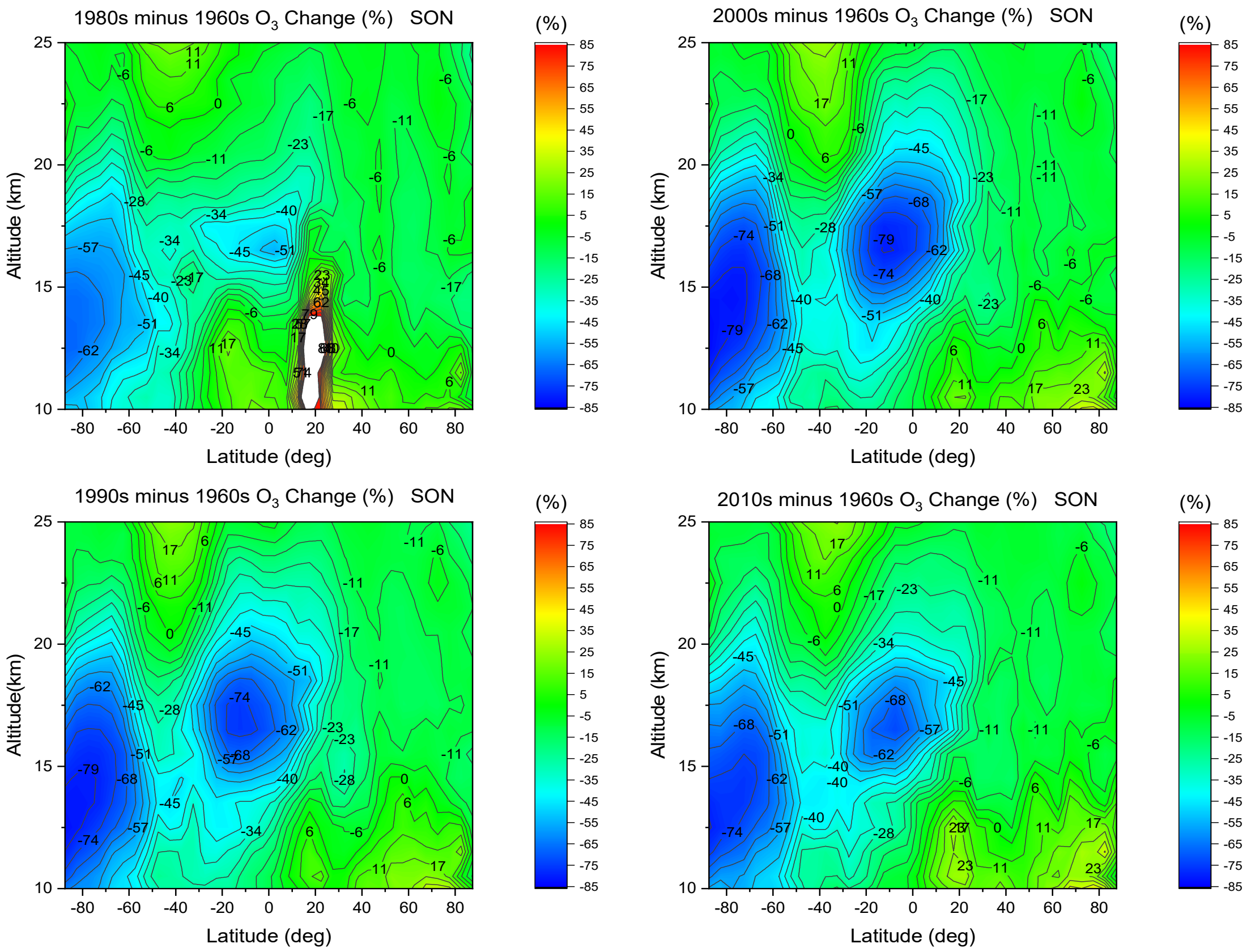

**Fig. 1**. Changes in % of the $O_3$ climatology in the season SON of the 1980s-2010s relative to that in the 1960s.

To show the relative $O_3$ changes in the Antarctic, tropical and Arctic stratospheres over the seasons, the $O_3$ climatology differences of the 2000s with respect to the 1960s for SON, DJF (December, January and February), MAM (March, April and May), and JJA (June, July and August) are plotted in Fig. 2. As expected, the polar $O_3$ holes have strong seasonal variations and mainly appear in their respective springs (SON for the Antarctica and MAM for the Arctic), whereas the tropical $O_3$ hole shows a weak sensitivity to seasons, existing in all seasons. The Arctic $O_3$ hole in MAM with a maximum $O_3$ loss of ~25% is far less and smaller in both depth and area than the Antarctic hole in SON or the tropical hole in any season. This is not unexpected, as the winter stratosphere of the Arctic is well known to be much warmer than that of the Antarctic or tropics and the Arctic vortex is not as well formed and as persistent as in the Antarctic or tropics.

Moreover, to show the robustness of the observations in Figs. 1-2, the relative seasonal $O_3$ changes in percent in 2000s are plotted with respect to those in the 1980s in Fig. 3. The latter solidly confirms the observation of the large and all-season ozone hole over the tropics, which was even slightly deeper than the springtime Antarctic ozone hole in SON (−62% versus −49%). In either the tropical or Antarctic ozone hole, ozone losses in Fig. 3 are smaller than those in Fig. 2. This is reasonable and expected, as ozone depletion was well known to become significant since 1980 (the beginning of the 1980s) and therefore there was a significant loss of ozone in average from the 1960s to the 1980s. The plotted results in Fig. 3 clearly confirm the validity of the data presented in Figs. 1-2 using the 1960s' data as the undisturbed ozone climatology.

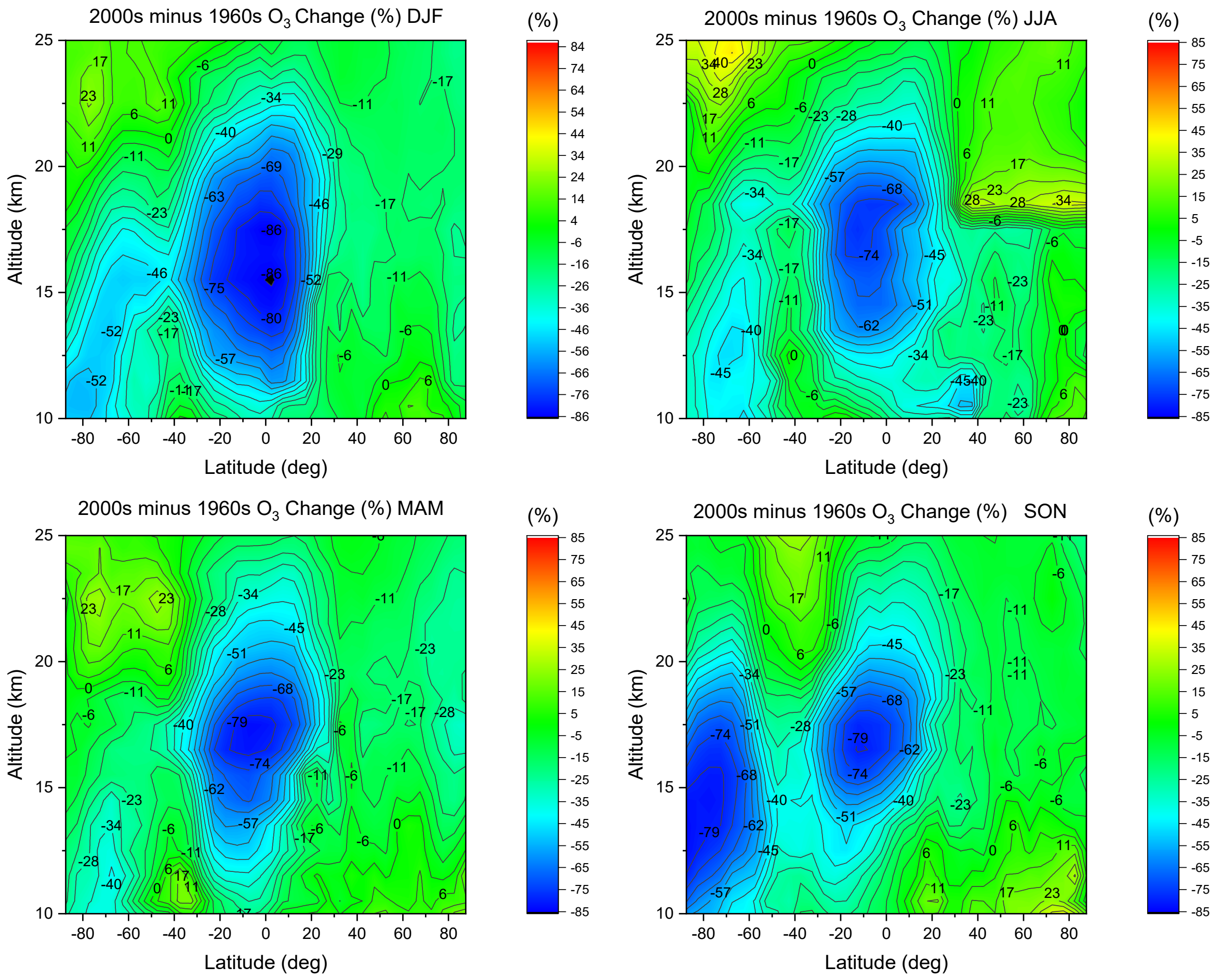

**Fig. 2**. Changes in % of the $O_3$ climatology in the seasons of the 2000s relative to those in the 1960s.

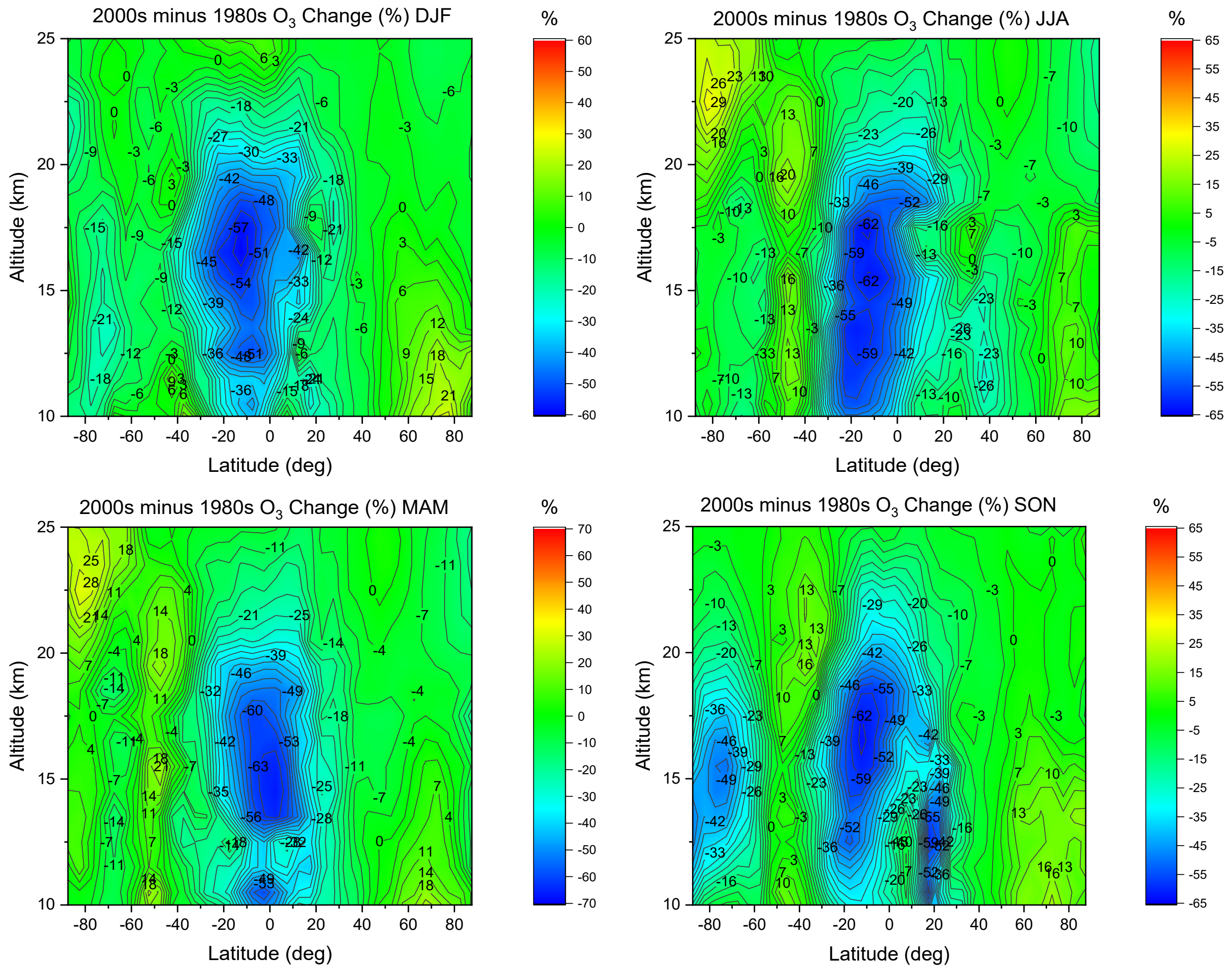

**Fig. 3**. Changes in % of the $O_3$ climatology in the seasons of the 2000s relative to those in the 1980s.

To show time-series depths of the decadal mean Antarctic, Arctic and tropical $O_3$ holes in SON, MAM and annual respectively, averaged $O_3$ changes in % at the lower stratospheric altitudes of 14-21 km are shown in Figs. 4a-c, which also include the data for the sum of concentrations of main ODSs (CFCs and $CCl_4$) measured in the troposphere (no transport lag times considered). It is interesting to show quantitatively that the Antarctic $O_3$ hole reached its maximum $O_3$ loss by about 60% in the 1990s, whereas the tropical hole reached its maximum loss by about 56% in the 2000s. By contrast, the Arctic $O_3$ hole reached its maximum $O_3$ loss by 20-25% in the 1990s, less than the half of the maximum $O_3$ loss in the Antarctic or tropical hole. Both tropical and polar ozone hole trends closely follow that of the total level of ODSs, having significant $O_3$ decreases in the 1980s and 1990s only with no significant trends over the past ~25 years.

Similar to our results shown in Fig. 4b, Polvani et al.[26] made an interesting observation. They observed large tropical ozone loss by about 300 ppbv from 1979 to 1997, though they only showed ozone anomalies (not $O_3$ loss in percentage) relative to the post-1998 mean at the single altitude of 18.5 km or at the pressure of 67 or 68 hPa over the tropics (30°S-30°N) from 1979 to 2013. They showed such a large tropical $O_3$ loss, consistently from three gridded datasets: the TOST for the period 1979-2013 (the same dataset as used for the present study),[28] the binary database of ozone profiles (BDBP) based on a compilation of observations obtained from several sources of high vertical resolution ozone profile measurements for the period 1979–2006,[29] and the NASA's global ozone chemistry and related trace gas data records for the stratosphere (GOZCARDS) merged satellite dataset based on a compilation of data from a number of satellite measurements for the period 1979–2012.[30] For a comparison, the results of the present study and Polvani et al.[26] are co-plotted in Figs. 5a-b. It is evidently shown that these observed data from multiple datasets of ground-based and combined ground-based and satellite measurements exhibit excellent agreements.

It is also worthwhile noting that the results in Fig. 4b indicate that no ozone hole over the tropics would be observed by the conventional definition of an ozone hole, as the total ozone over the tropics were above 220 DU even with a loss by 56% of the column ozone at altitudes of 14-21 km, given that the undisturbed total ozone was about 270 DU over the tropics, of which only approximately 25% is distributed at 14-21 km. In contrast, with the new definition of an ozone hole in the present study, a pronounced tropical $O_3$ hole is observed, as clearly seen in Figs. 1-4.

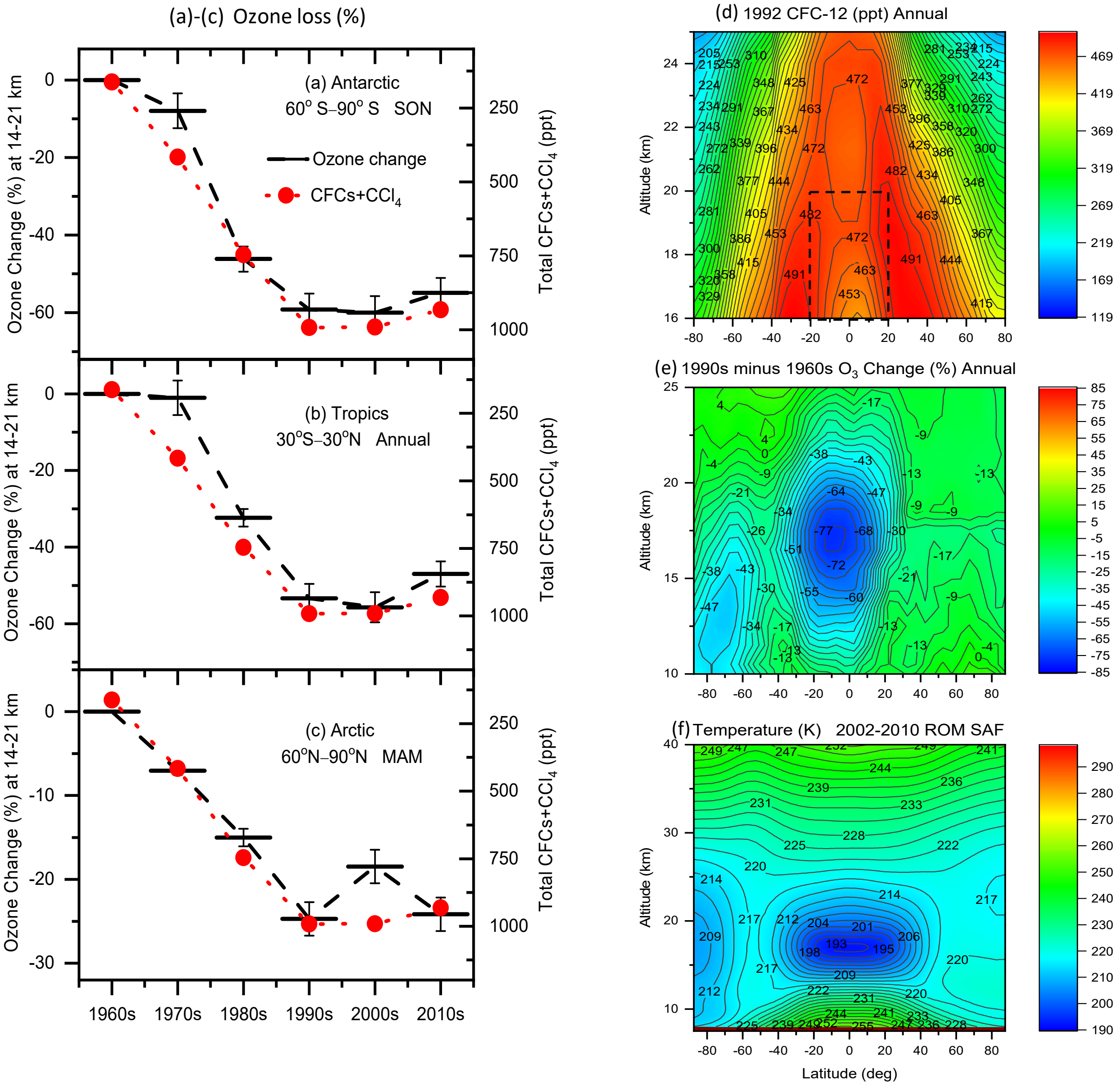

**Fig. 4**. a-c: Time-series decadal mean $O_3$ changes in % with respect to that of the 1960s at the lower stratosphere of 14-21 km of the Antarctic in SON, the Arctic in MAM and the annual tropical (solid lines in black) and sum of measured concentrations of main ODSs (CFCs and $CCl_4$) (solid circles in red) over the 1960s-2010s period; d: Zonal mean latitude-altitude distribution of the $CF_2Cl_2$ (CFC-12) concentration in 1992 obtained from the NASA UARS's CLEAS dataset, where the dash square shows the zone of most significant CFC destruction; e: the difference in % of the annual $O_3$ climatology in the 1990s with respect to the 1960s; f: Decadal mean zonal mean latitude-altitude distribution of the temperature climatology averaged over seasons in the 2000s.

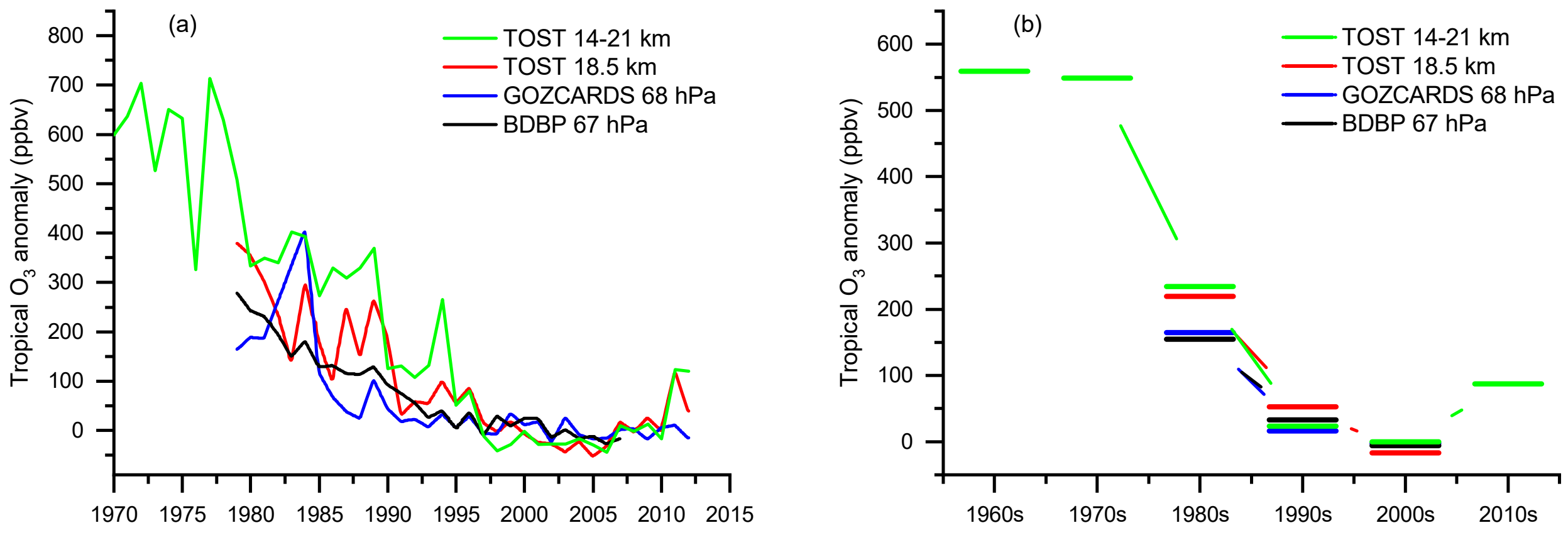

**Fig. 5**. Comparison of time-series annual and decadal mean tropical (30°S–30°N) ozone anomalies at the lower stratosphere of 14-21 km relative to the 2000s mean from the TOST[28] (the present study) with those at the altitude of 18.5 km or 67/68 hPa relative to the post-1998 mean from three gridded datasets by Polvani et al.[26]: the TOST for the period 1979–2013,[28] the binary database of ozone profiles (BDBP) for the period 1979–2007,[29] and the NASA's global ozone chemistry and related trace gas data records for the stratosphere (GOZCARDS) for the period 1979–2012.[30]

### 3.2 CFC-12 distribution and tropical $O_3$ hole

The author[6-8] has previously shown that significant decompositions of CFCs and $N_2O$ but not $CH_4$ occur in the lower Antarctic stratosphere during winter. The data provided solid evidence of the CRE reactions of CFCs and $N_2O$, but not $CH_4$, in the *winter* polar stratosphere, though $CH_4$ is not even a secure 'inert trace gas' because it can also react with radicals generated from the reactions of CFCs or $N_2O$. Here zonal mean latitude-altitude distribution of $CF_2Cl_2$ (CFC-12) in 1992 from the dataset of the NASA UARS (CLEAS) and the relative difference averaged over the seasons of the $O_3$ climatology in the 1990s with respect to the 1960s are respectively plotted in Figs. 4d-e. The CFC-12 images over the four seasons are shown in Fig. S3. The data clearly show the depletion of CFC-12 in the winter and early spring lower polar stratosphere, as observed previously[6-8]. Most importantly, the data in Figs. 4d and S3 show that in all seasons, the CFC-12 concentration was depleted in the lower stratosphere below 25 km over the tropics (at latitudes of 30°S-30°N), most significant in the zone at 16-20 km and at 20°S-20°N, in which correspondingly the circularly symmetric annual mean tropical $O_3$ hole is centrally located (Fig. 4e).

### 3.3 Time-series total ozone and Umkehr datasets

Time-series total $O_3$ data at the springtime Antarctic (60°-90°S) and the annual tropics (30°S-30°N) measured by satellites since the 1979 are shown in Figs. 6a-b. It is seen that the decreases in absolute total $O_3$ values over the tropics are smaller than those over the Antarctic. This is expected due to the difference between the undisturbed $O_3$ distributions over the Antarctic and the tropics (as mentioned above). The 11-year cyclic variation of total $O_3$ over Antarctica has been well reproduced by the CRE equation previously.[6-8] Here, we use the CRE Eq. (1) to give a quantitative analysis of observed total $O_3$ data. Given the observed distributions of the CR ionization rate and CFCs versus latitude and the use of the CR intensity measured at the Antarctic and the concentrations measured at the lower troposphere of CFCs which are well mixed globally as inputs to calculate $O_3$ loss over both the Antarctic and tropics, the reaction constant $k$ in Eq. (1) for the Antarctic is approximately double that for the tropics, i.e., $k_{Tro}\approx½k_{Ant}$. Another factor must also be considered: the transport lag times of CFCs from the troposphere to the lower stratosphere over the Antarctic and the tropics, which are about 1 year and 10 years respectively determined from analyses of observed data in terms of the CRE mechanism.[6-9] Moreover, we make a rough assumption that only the $O_3$ layer in the lower tropical stratosphere below 25 km (i.e, ~25% of the total $O_3$) is affected by the CRE mechanism. Taking all these factors into account, the CRE-modeled results are shown in Figs. 6a-b. It is seen that given the fluctuation level of observed data, the results fitted by Eq. (1) for the Antarctic $O_3$ hole are excellent. This gives rise to the best fitting constant $k_{Ant}$, half value of which ($k_{Tro}\approx½k_{Ant}$) is used to calculate the $O_3$ loss over the tropics *with no adjustments*. We notice that the modeled results for tropical $O_3$ loss is not as good as for the Antarctic. Given several rough approximations made in this modeling, however, the modeled results reproduce the observed $O_3$ trends surprizingly well and are able to capture the main characteristic of 11-year cyclic variations.

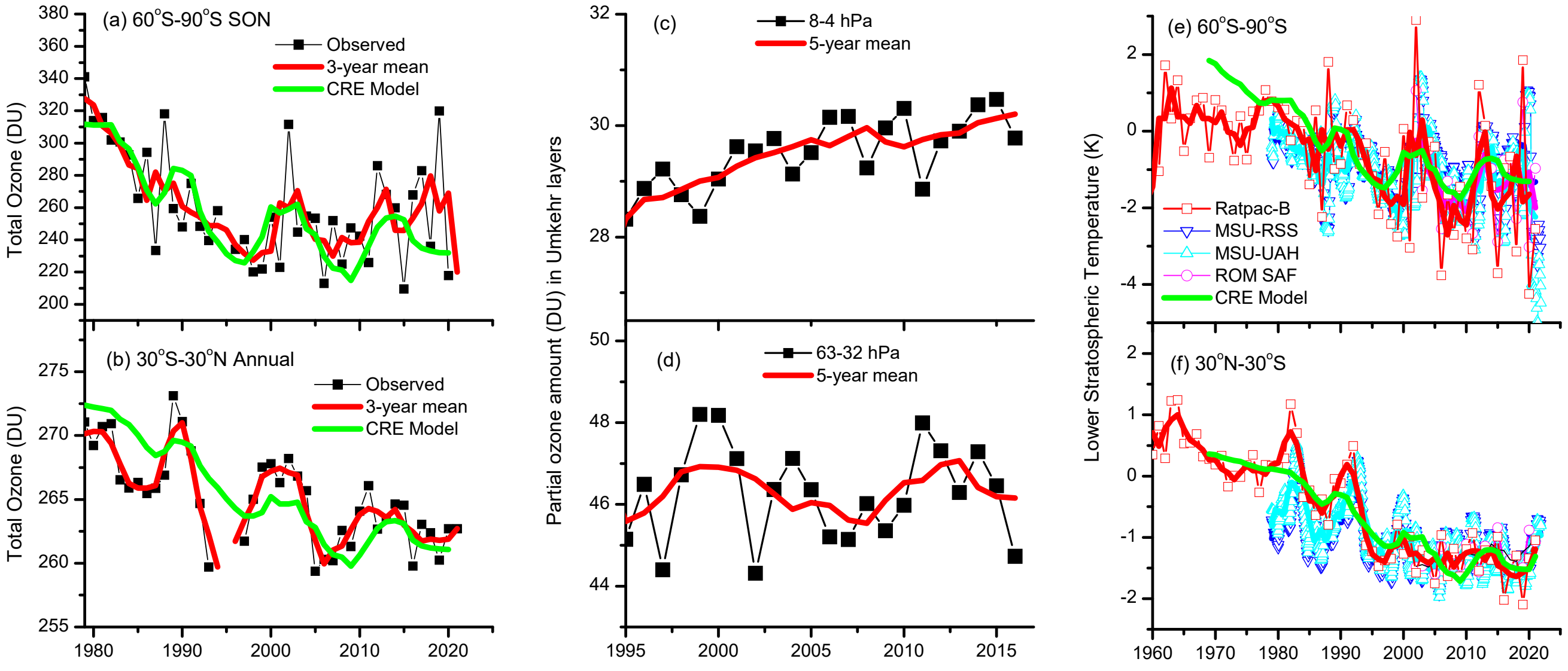

**Fig. 6**. a-b: Time-series total ozone at the springtime Antarctic (60°-90°S) and the annual tropics (30°S-30°N) measured by NASA TOMS, OMI and OMPS satellites since the 1979. c-d: Time-series Umkehr data of annul mean vertical distribution partial column ozone at the lower and upper stratospheric layers (63-32 hPa and 8-4 hPa) at Mauna Loa (19.5°N, 155.6°W, HI, USA). e-f: Time-series annul mean lower stratospheric temperature anomalies of the Antarctic and the tropics since 1960, obtained from multiple ground- and satellite-based data measurements (Ratpac, MSU-UAH, MSU-RSS and ROM SAF) and offset to compare with each other. Also shown are the fitted total $O_3$ and temperature curves given by the CRE equation (see text), as well as 3(5)-year smoothing (thick solid lines in colors) to observed data (symbols).

Moreover, the Umkehr dataset for the Mauna Loa (19.5°N, 155.6°W, HI, USA) is shown in Figs. 6c-d. For the Antarctic $O_3$ hole, the 11-year cyclic oscillation of total $O_3$ was mainly due to the CRE-caused $O_3$ loss in the stratospheric layers at which the CR ionization intensity was strong, with the most pronounced effect at the Umkehr layer of 63-32 hPa (21-25 km) slightly above the CR-peak layer of 126-63 hPa (15-21 km) in which the CRE mechanism overkilled the $O_3$ layer.[9] In contrast, no 11-year cyclic variations were observed for $O_3$ in the upper stratospheric layer where the photochemical mechanism dominates.[9, 31] A similar behavior is now observed for $O_3$ loss over the tropics, as shown in Figs. 6c-d. The results in Figs. 4a-b and 6a-d indicate that $O_3$ depletion over the Antarctica and the tropics must arise from an identical physical mechanism.

3.4 Lower stratospheric temperature (LST)

It is well known that $O_3$ depletion causes stratospheric cooling; temperature drop in the lower stratosphere is a direct measure of $O_3$ loss. The seasonal zonal mean latitude-altitude distribution of the temperature at altitudes of 8-40 km for 2002-2010, which is made from the zonally averaged monthly means on a latitude-altitude grid of radio occultation (RO) satellite datasets since 2002, obtained from the high-quality ROM SAF's Climate Data Record (CDR) datasets,[32] is shown in Fig. S4. Remarkably, the data show that the temperature in the low-temperature cyclones over the tropics over the seasons is 190-200 K, being only 5-8 K above that in the Antarctic polar vortex in winter (JJA) but 13-16 K below that in the Arctic polar vortex in winter (DJF). This can well explain why the $O_3$ hole over the Antarctic or tropics is much larger and deeper than the Arctic $O_3$ hole and why the tropical $O_3$ hole is constantly formed over the seasons. The annul mean zonal mean latitude-altitude distribution of the temperature averaged over the four seasons is shown in Fig. 4f. The latter shows a remarkable phenomenon that the annul mean LST over the tropics is the lowest at 193-198 K, compared with 209-217 K over the Antarctic and 217-220 K over the Arctic. As shown in Fig. 4e, correspondingly the annul mean $O_3$ depletion is the largest in the tropical $O_3$ hole, which is 77% versus 47% for the Antarctic and versus ~10% for the Arctic in the 1990s when the polar holes peaked. Consistent with the observation from ozone datasets shown in Figs. 1-5, the stratospheric temperature data in Figs. 4f and S4 exhibit a pronounced temperature 'hole' over the tropics, providing strong evidence of a deep and large tropical $O_3$ hole, along with the well-known Antarctic and Arctic $O_3$ holes.

Furthermore, the author has well demonstrated that over the Antarctic, the LST regulated by stratospheric cooling due to $O_3$ loss has pronounced 11-year cyclic variations and can be well reproduced by the CRE equation.[6-9] Here, Figs. 6e-f show time-series temperature anomaly datasets at the lower stratospheres (100-30 hPa) over the Antarctic and the tropics from multiple satellite and ground-based measurements, including NOAA's MSU UAH[33] and RSS[34] satellite datasets, ROM SAF's RO satellite datasets[32], and NOAA's radiosonde-based Ratpac-B time-series dataset[35]. Note that the LST anomalies in original datasets were normalized to various reference temperatures and had therefore to be offset to compare with each other, but this offsetting has no effects on their long-term trends. These datasets for the Antarctic solidly confirm 11-year cyclic LST variations reported previously,[6-9] while the LST over the tropics shows similar cyclic variations. Moreover, all the datasets show that *significant $O_3$ or LST reductions only occurred in the 1980s and 1990s with no significant trends over the past ~25 years*. This is similar to the observation by Polvani et al[26] and those reported in the newest IPCC AR6[36]. The latter states: "most datasets show that lower stratospheric temperatures have stabilized since the mid-1990s with no significant change over the last 20 years" (see page 2-48, chap. 2 of the Report[36]). Using the same relationship of $k_{Tro}≈½k_{Ant}$ as that used for modeling the $O_3$ data in Figs. 6a-b, we now show that the CRE equation fits to the observed LST data in Figs. 6e-f reasonably well. Again, these results clearly confirm that the Antarctic and tropical $O_3$ holes must arise from an identical physical mechanism, for which the CRE mechanism has given the best agreements with observations. So far, the CRE mechanism is the only physical mechanism that has well reproduced the observed 11-year cyclic stratospheric $O_3$ and LST variations.[4-9]

## 4. DISCUSSION

The present finding of a tropical ozone hole is closet to the observation by Polvani et al.[26] of large $O_3$ losses at the altitude of 18.5 km or 67/68 hPa over the tropics (30°S-30°N) from 1979 to 1997 with data from three independent datasets (TOST, BDBP and GOZCARDS)[28-30], though they did not reveal the tropical $O_3$ hole. Polvani et al.[26] also showed that global and tropical LST cooling had disappeared since 1997, while the tropical $O_3$ concentration reached the minimum around 2005. In explaining their observed results, however, Polvani et al. argued for the low abundance of active chlorine and hence no local chemical $O_3$ destruction in the tropical lower stratosphere. Instead, they argued that the observed tropical $O_3$ and cooling trends were primarily driven by *tropical upwelling* caused by ODSs rather than greenhouse gases (GHGs) (mainly non-halogenated GHGs, as widely believed). As we have noted in Introduction and will further discuss later, however, even a low level of active halogen can cause significant ozone depletion in the tropical lower stratosphere.

Interestingly, Polvani et al.[26] also performed simulations from a chemistry–climate model (CCM) with incrementally added single forcings (sea surface temperatures—SSTs, GHGs, ODSs, volcanic eruptions and solar variations) to detail the contribution of each forcing to tropical ozone and LSTs. Although their simulated results showed that ODSs were the dominant forcing of tropical ozone loss over GHGs, it must be pointed out that their simulated values of sum ozone loss (−28±13 ppbv per decade, see their Table 1) were about *five times* smaller than their observed results (~−150 ppbv per decade for the 1980 and 1990s), even ignoring that not all individual ensemble members showed statistically significant trends (see their Table 2). Moreover, in contrast to their claim that tropical lower stratospheric ozone would be closely tied to tropical upwelling w*, their simulated value of the w* increase by ODSs is *not* dominant but very close to that by GHGs, each force contributing to an increment by ~0.1 km $yr^{-1}$ $decade^{-1}$ at 85 hPa, namely 0.04±0.09 for SSTs, 0.10±0.11 for SSTs+GHGs and 0.21±0.11 for SSTs+GHGs+ODSs (see their Table 1). The latter results were also consistent with their simulated results of tropical LST trends (see their Fig. 3).

The simulated results of CCMs by Randel and Thompson[37] and others[38] had some differences from but were overall similar to the above-mentioned CCM results by Polvani et al.[26] In most simulated results of CCMs, the strength of tropical upwelling was projected to increase from 1960 to 2100 by *~2% per decade* with the largest trends occurring in JJA, corresponding to tropical $O_3$ reductions at 50 hPa of 0.15-0.35 ppmv (*11-25 ppbv per decade*)[38]. This ozone loss trend resulting from CCM simulations is *about 10 times less than* the observations by Polvani et al.[26] and the present observations shown in Figs. 1-5. More crucially, the observed data have robustly shown that significant tropical ozone loss and LST cooling occurred *in the 1980s and 1990s only*, which is in drastic discrepancy from the simulated results of CCMs. Polvani et al.[26] was then led to the open question: How could ODSs affect the stratospheric circulation? They conceded that the underlying mechanism for ODSs being a key forcing for tropical lower stratospheric $O_3$ and temperature trends remained largely unexplored. Knowing the wide belief in CCMs that the key drivers of tropical upwelling and thus tropical $O_3$ or LST trends since 1960 are non-halogenated GHGs (mainly $CO_2$), Polvani et al.[26] were forced to suggest that polar $O_3$ depletion caused by ODSs would cause tropical upwelling and hence large tropical $O_3$ losses. This explanation cannot be correct either, as the polar $O_3$ hole is seasonal and appears only in the springtime, whereas the tropical $O_3$ hole is all-season and has no changes in its central location over the seasons and over the decades since its appearance in the 1980s (Figs. 1-3 and 4e).

The present observed results in Figs. 1-6 and S1-S4 strongly indicate that, like the Antarctic $O_3$ hole that was once incorrectly explained by the misconceived air transport mechanism ('dynamical theory'), the tropical $O_3$ hole must not result from changes in normal atmospheric circulation patterns over the tropics since the 1960s or 1970s but result from an identical physical/chemical mechanism to that for the polar $O_3$ hole. Obviously the tropical $O_3$ hole varies closely with the atmospheric level of CFCs (as seen in Fig. 4b), so it must originate from a CFC-related mechanism. The postulated stratospheric cooling and tropical upwelling effects of increasing non-halogenated GHGs have disappeared in observed $O_3$ and temperature data for the Antarctic lower stratosphere[6-8] and for the tropical lower stratosphere[26]. It is obvious that the simulated results from CCMs[26, 37, 38] do not agree with the observed results shown in Figs. 4b, 5a-b and 6f, which show that the negative $O_3$ trends were about 10 times larger (*−25 to −30% per decade*) in the 1980s and 1990s and there have been *no significant $O_3$ or LST trends* in the tropical/Antarctic since the mid-1990s. The latter is actually consistent with the observations summarized in the newest IPCC Report[36]. Moreover, the proposed *enhanced* tropical upwelling directly contradicts with the observed CFC depletion in the lower tropical stratosphere (Fig. 4d), as increased upward motion would transport CFC-rich air from the troposphere. Indeed, the observed data robustly show no shifts in the positions of both Antarctic and tropical $O_3$ holes that have constantly been centered at the altitude region corresponding to the CR ionization peak since the 1960s/1980s and circularly symmetric $O_3$ depletion cyclones are formed with the largest depletion at the centers (Fig. 1-3 and 4e). These major features cannot be explained by tropical upwelling due to non-halogenated GHGs (mainly $CO_2$) that have kept rising since the 1960s. All the observed data strongly indicate that tropical upwelling cannot be the major mechanism for the observed large, deep and all-season tropical $O_3$ hole. The simultaneous depletions of both CFCs and $O_3$ in the lower tropical stratosphere are most likely due to a physical reaction mechanism that occurs locally. For the latter, the CRE mechanism, supported by the observed data in Figs. 1-6 as well as the substantial datasets obtained from both laboratory and atmospheric measurements[3-9, 39-41], has provided the best and predictive model.

It is well known that the presence of PSCs is crucial for formation of the Antarctic $O_3$ hole.[42-45] It was proposed that on the surfaces of PSCs, chlorine reservoir molecules (HCl and $ClONO_2$) are converted into photoactive forms ($Cl_2$) that can then undergo photolysis to destroy $O_3$. There are two types of PSCs, namely Type I and Type II PSC. The composition of Type II PSC is water ice, while Type I PSC is composed of mixtures of nitric acid ($HNO_3$), water vapor ($H_2O$), and sulfuric acid ($H_2SO_4$). The temperatures required for formation of Type I and II PSCs are 195K and 188K, respectively. Thus, it is very likely that TSCs, at least Type I PSC-like TSCs, can also form in the tropical lower stratosphere over the seasons due to the observed low temperatures of 190-200 K (Figs. S4 and 4f). CRs may also play a certain role in forming PSCs and PSC-like TSCs.[16, 19] Note that the tropical lower stratosphere is very different from the polar lower stratosphere in both composition and climate. The former is rich of CFCs and other halogen-containing gases, whereas the latter is composed of inorganic chlorine species and lower-level CFCs. However, the CRE mechanism put forward two decades ago has proposed that $O_3$-depleting reactions of both CFCs and inorganic halogen species can effectively occur on the surfaces of PSCs.[3-9, 39-41] Therefore, there are required and sufficient conditions for $O_3$-depleting reactions to occur on the surfaces of proposed TSCs in the tropical lower stratosphere. As noted in Introduction, the constant co-presence of low-temperature TSCs and intense sunlight should lead to a unique active halogen evolution in the tropical lower stratosphere, in which halogen-catalyzed reactions are much more efficient for $O_3$ destruction than that in the polar lower stratosphere.

The tropics (30°N-30°S) constitutes 50% of Earth's surface area, which is home to about 50% of the world's population. $O_3$ depletion in the tropics could cause a large global concern. In areas where $O_3$ depletion is observed to be smaller in absolute $O_3$ value, UV-B increases are more difficult to detect as the detection can be complicated by changes in cloudiness, local pollution, and other difficulties. However, it is generally agreed that the depletion of the $O_3$ layer leads to an increase in ground-level UV radiation because ozone is an effective absorber of solar UV radiation. Exposure to enhanced UV-B levels could increase the incidence of skin cancer and cataracts in humans, weaken human immune systems, decrease agricultural productivity, and negatively affect sensitive aquatic organisms and ecosystems[46]. Indeed, there was a report called HIPERION published by the Ecuadorian Space Agency in 2008[47]. The study using ground measurements in Ecuador and satellite data for several countries over 28 years found that the UV radiation reaching equatorial latitudes was far greater than expected, with the UV Index as high as 24 in Quito. This Ecuadorian report concluded that $O_3$ depletion levels over equatorial regions are already endangering large populations in the regions. Further delicate studies of $O_3$ depletion, UV radiation change, increased cancer risks and other negative effects on health and ecosystems in the tropical regions will be of great interest and significance.

Another important result is that the global lower stratospheric temperature is essentially governed by the $O_3$ layer, which is expected as ozone is the main and dominant molecule that absorbs solar radiation in the stratosphere. As a result, the presence of the tropical and polar $O_3$ holes will play a major role in stratospheric cooling and regulating the global lower stratospheric temperature, as seen previously[6-9, 26] and in the results shown in Figs. S4, 4f, and 6e-f. As seen in Figs. S4 and 4f, this is equivalent to the formation of three temperature 'holes' in the stratosphere, corresponding to the Antarctic, tropical and Arctic $O_3$ holes respectively. This result will be further explored in a subsequent paper.

## 5. CONCLUSIONS

In summary, a critical review of the CRE mechanism led to a hypothesis of an ozone hole in the lower stratosphere over the tropics. Substantial datasets of troposphere-stratosphere ozone climatology since the 1960s, latitude-altitude CFC distribution, time-series total ozone, Ozonesonde and Umkehr datasets, and lower stratospheric temperature climatology from multiple ground-based or satellite measurements have provided strong evidence of the existing whole-year ozone hole over the tropics. At the hole center, up to approximately 80% of the ozone concentrations relative to the normal value in the 1960s is depleted. The averaged depth of the annual tropic ozone hole is comparable to that of the springtime Antarctic ozone hole, while the size of the tropic hole is about seven times that of the Antarctic hole. The annul mean ozone depletion in the lower stratosphere over the tropics is strikingly large, which is 77% versus 47% over the Antarctic and versus ~10% over the Arctic in the 1990s when the polar holes were in their peaks. The observed results strongly indicate that both the Antarctic ozone hole and the tropical ozone hole must arise from an identical physical mechanism and the CRE mechanism has provided the best predictive model which exhibits good agreements with observations. The tropical lower stratosphere is rich of CFCs and other halogenated ODSs and CR ionization events, and critically its temperature for all seasons is sufficiently cold for formation of PSC-like TSCs. Therefore, there is every reason to conclude that the CRE mechanism of ozone-depleting reactions have most likely taken place on the surfaces of TSCs present in the lower stratosphere over the tropics, leading to the formation of the tropical ozone hole in the whole year.

## SUPPLEMENTARY MATERIAL

See supplementary material for Figs. S1-S4.

## ACKNOWLEDGEMENTS

The author is greatly indebted to the Science Teams for making the data used for this study available. This work is supported by the Natural Science and Engineering Research Council of Canada.

## AUTHOR DECLARATIONS

The author has no conflicts to disclose.

## DATA AVAILABILITY

The data used for this study were obtained from the following sources: the TOST datasets were obtained from the WMO's World Ozone and Ultraviolet Radiation Data Centre (WOUDC) (https://woudc.org/data.php); zonal mean latitude-altitude distribution of $CF_2Cl_2$ (CFC-12) in 1992 was obtained from the NASA UARS (CLEAS) dataset (https://earthdata.nasa.gov/ or https://uars.gsfc.nasa.gov/Public/Analysis/UARS/urap/home.html); time-series total ozone data at the springtime Antarctic (60°-90°S) and the annual tropics (30°S-30°N) were obtained from NASA TOMS, OMI and OMPS satellite datasets (https://ozonewatch.gsfc.nasa.gov/meteorology/SH.html or https://woudc.org/data.php); the Umkehr dataset of ozone for the Mauna Loa (19.5°N, 155.6°W, HI, USA) was obtained from NOAA's Global Monitoring Laboratory (https://gml.noaa.gov/aftp/data/ozwv/) and WOUDC (https://woudc.org/data.php); lower stratospheric temperature data of NOAA's MSU UAH and RSS datasets and Ratpac-B time-series dataset were obtained from NOAA (https://www.ncdc.noaa.gov/climate-monitoring/) and of RO satellite datasets were obtained from the ROM SAF (https://www.romsaf.org/product_archive.php); cosmic ray data were obtained from ref.[9]; tropospheric data of CFCs and $CCl_4$ were obtained from the 2021 IPCC AR6 Report[36].